\documentclass[aps,prd,preprintnumbers,showpacs,superscriptaddress,nofootinbib,amsmath,amssymb,floats,floatfix,showkeys,notitlepage,longbibliography]{revtex4-1}
\usepackage{comment}
\usepackage{graphicx}
\usepackage{subfigure}
\usepackage{palatino}
\usepackage[commandnameprefix=always]{changes}
\usepackage{hyperref}
\hypersetup{colorlinks=true,linkcolor=red,urlcolor=red,citecolor=red}
\usepackage[toc,page]{appendix}
\usepackage[normalem]{ulem}

\usepackage{lipsum}
\usepackage{graphicx}
\usepackage{subfigure}
\usepackage{palatino}
\usepackage{float}
\usepackage{sans}
\usepackage{adjustbox}
\usepackage{latexsym}
\usepackage{amsmath}
\usepackage{amssymb}
\usepackage{amsfonts}
\usepackage{dcolumn}
\usepackage{bm}
\usepackage{tikz}
\usepackage{bigints}
\usepackage{array,tabularx,multirow,booktabs}
\usepackage[tracking=true]{microtype}
\SetTracking{}{500}
\SetTracking{encoding={*}, shape=sc}{40}
\allowdisplaybreaks
\usepackage{adjustbox}
\usepackage{latexsym}
\usepackage{amsmath}
\usepackage{amssymb}
\usepackage{amsfonts}
\usepackage{dcolumn}
\usepackage{bm}
\usepackage{tikz}
\usepackage{bigints}
\usepackage{array,tabularx,multirow,booktabs}
\usepackage[tracking=true]{microtype}
\usepackage{color}
\allowdisplaybreaks

\begin{document}

\title{
Topological Insights into Black Hole Thermodynamics: Non-Extensive Entropy in CFT framework}

\author{Mohammad Ali S. Afshar}
\email{m.a.s.afshar@gmail.com}
\affiliation{Department of Physics, Faculty of Basic
Sciences, University of Mazandaran\\ P. O. Box 47416-95447, Babolsar, Iran}

\author{Mohammad Reza Alipour}
\email{mr.alipour@stu.umz.ac.ir}
\affiliation{School of Physics, Damghan University, P. O. Box 3671641167, Damghan, Iran}
\affiliation{Department of Physics, Faculty of Basic
Sciences, University of Mazandaran\\ P. O. Box 47416-95447, Babolsar, Iran}

\author{Saeed Noori Gashti}
\email{saeed.noorigashti@stu.umz.ac.ir; saeed.noorigashti70@gmail.com}
\affiliation{School of Physics, Damghan University, P. O. Box 3671641167, Damghan, Iran}

\author{Jafar Sadeghi}
\email{pouriya@ipm.ir}
\affiliation{Department of Physics, Faculty of Basic
Sciences, University of Mazandaran\\ P. O. Box 47416-95447, Babolsar, Iran}

\begin{abstract}
In this paper, We conducted an in-depth investigation into the thermodynamic topology of Einstein-Gauss-Bonnet black holes within the framework of Conformal Field Theory (CFT), considering the implications of non-extensive entropy formulations. Our study reveals that the parameter $\lambda$ (R\'{e}nyi entropy) plays a crucial role in the phase behavior of black holes. Specifically, when $\lambda$ is below the critical value (C), it has a negligible impact on the phase behavior. However, when $\lambda$ exceeds the critical value, it significantly alters the phase transition outcomes. Determining the most physically representative values of $\lambda$ will require experimental validation, but this parameter flexibility allows researchers to better explain black hole phase transitions under varying physical conditions. Furthermore, the parameters $\alpha$ and $\beta$ affect the phase structure and topological charge for the Sharma-Mittal entropy. Only in the case of $C>C_c$ and in the condition of $\alpha\approx\beta$ will we have a first-order phase transition with topological charge + 1. Additionally, for the loop quantum gravity non-extensive entropy as the parameter $q$ approaches 1, the classification of topological charges changes. We observe configurations with one and three topological charges with respect to critical value $C$, resulting in a total topological charge $W = +1$, and configurations with two topological charges $(\omega = +1, -1)$, leading to a total topological charge $W = 0$. These findings provide new insights into the complex phase behavior and topological characteristics of black holes in the context of CFT and non-extensive entropy formulations.
\end{abstract}

\date{\today}



\maketitle
\tableofcontents
\section{Introduction}
The AdS/CFT correspondence is a powerful concept in theoretical physics that proposes a duality between two types of physical theories: gravitational theories in Anti-de Sitter (AdS) space and conformal field theories (CFT) defined on the boundary. This correspondence can open numerous avenues for diverse and varied studies in cosmological physics.
The AdS/CFT correspondence provides a rich framework for exploring the connections between quantum field theories, thermal dynamics, and gravity. In this context, the thermodynamic properties of black holes in AdS space (such as temperature and entropy) have direct analogs in the thermodynamics of CFT, deepening our understanding of the fundamental laws of the universe. For instance, black holes residing in AdS space exhibit Hawking radiation (thermal particles emitted from black holes), which can be understood based on the dual degrees of freedom in CFT. The emitted radiation corresponds to excitations in the boundary theory, indicating a profound connection between black hole dynamics and field theory \cite{1,2,2',3}. Conversely, the CFT living on the boundary of AdS space has its own thermodynamic characteristics. In many cases, especially at high temperatures, this CFT can be described using a statistical ensemble, where states are thermally populated according to their energy. Consequently, the CFT partition function can be computed via the Euclidean path integral on the boundary. The free energy of the CFT can be related to the gravitational action computed in the bulk AdS space. Establishing such bridges allows for the calculation of various thermodynamic observables, such as Helmholtz free energy, Gibbs free energy, and entropy, from black hole solutions in AdS and the corresponding CFT \cite{4}. Additionally, the rich phase structure in AdS black holes enables the analysis of phase transitions between different thermodynamic states, linking them to the structural properties of the corresponding CFT. In summary, when a black hole forms in AdS, the corresponding thermal state in the CFT can be identified.\\
One of the most intriguing and significant aspects of classical thermodynamics, which typically contains vital information about the behavior of the sample under study, is the examination of phase transitions. Phase transitions are traditionally analyzed by examining the behavior of temperature-dependent plots of the Gibbs free energy. In a first-order phase transition, this results in the appearance of a swallowtail form, while in a second-order phase transition, it leads to a smoothly curved plot.
The use of this traditional method to study the thermodynamics of black holes has been explored in various researches\cite{500,501,502}. However, it must be acknowledged that due to the complexity of black hole field equations, these studies, particularly for the swallowtail form, have encountered significant difficulties.\\
In 2017, Cunha and colleagues \cite{5,5'} proposed using the extrema of the rewritten effective potential, combined with the concept of winding, to study the topology of photon rings and photon spheres in flat black holes. This method was subsequently extended by Wei to asymptotically AdS and dS black holes \cite{6}, leading to numerous studies based on this approach and relative validation of the method \cite{7,7',8,8',8'',8''',9,9'}.
Since the overall method was based on the structure of the scalar quantity of energy, it suggested that scalar functions exhibiting similar behavior could benefit from this general method for study. Accordingly, Wei extended this method to thermodynamics and, for the first time, applied it to the study of critical points of a sample based on the behavior of the temperature function\cite{10}. This preliminary work primarily focused on finding the critical points of black holes as a thermodynamic sample, and did not provide much information about phase transitions.
Subsequently, Wei and colleagues extended this method to the Helmholtz free energy \cite{11}, which in initial studies indicated very interesting information about black hole phase transitions. Following this innovation, numerous studies were conducted in both forms, particularly the free energy method, demonstrating its effectiveness \cite{12,12',13,13',13'',14,14',14'',15,15',15'',16,16',17,17',18,18',18'',19,19',20,20',20'',21,21',21'',21''',21'''',22,22',22'',22''',300,301,302,303,304,305,306,307}. In fact, the F method in these studies showed that it could effectively represent the phase transition behavior of black holes.\\
Given this background, it is essential to clearly state our motivation and objective in this study. In the method of using the Helmholtz free energy (F) to study phase transitions, whether in the traditional form or the topological method, we consider a thermodynamic parameter as the target parameter for classifying the phase transition space. This is based on the first law of thermodynamics and the canonical perspective of the sample. We then examine the phase behavior of the sample around the critical points of this parameter. Typically, this target parameter is temperature or pressure.
In this paper, we aim to shift the role of the target parameter to the central charge and investigate the phase behavior of the sample around the critical points of this parameter, to see if this difference will lead to a different view of the sample's topological phase transition.
Also, since the definition of free energy is intertwined with the concept of entropy, we will not limit our study to the usual Bekenstein-Hawking entropy. Instead, we will incorporate super statistical entropies to examine and compare the impact of additional parameters of these new entropies on phase transitions.
Regarding the use of super statistical entropies in this research, it should be noted that although comprehensive studies on the mutual effects of super statistics and gravitational structures have not yet been systematically conducted, and most studies are scattered, these studies show that the need for non-standard and super statistical descriptions is one of the essential requirements for the development of studies. \cite{23,23'}.
Traditional statistical mechanics typically relies on invoking equilibrium distributions (such as the Boltzmann distribution). However, many real-world systems are inherently non-equilibrium, influenced by varying external conditions or local interactions that can lead to significant fluctuations in energy, temperature, or other thermodynamic variables. Super statistics provides a framework for describing such systems by incorporating local statistical variations.
Therefore, it seems essential to pay more attention to the contribution and impact of super statistical perspectives in studies of systems with complex behaviors, such as black holes.\\
With respect to all the above statements, we arrange the paper as follows. In section II, we briefly review the model. In Sections III and IV we study the overview of the thermodynamic topology method and introduction of nonextensive entropy such as R\'{e}nyi, Sharma-Mittal, and loop quantum gravity. In Section V, VI, VII we will study and analyze using the different superstatistical entropies for the introduced model Finally, we have conclusions which are summarized in Section VIII.
\section{Gauss-Bonnet-AdS black hole}
The Gauss-Bonnet-AdS black hole is studied in detail. The action for the d-dimensional Einstein-Maxwell theory is presented, incorporating the Gauss-Bonnet term along with a negative cosmological constant \cite{24,25,26,27},
\begin{equation}\label{eq1}
S=\frac{1}{16\pi G}\int d^dx \sqrt{-g}\bigg[R-2\Lambda+\alpha_{GB}(R_{\mu\nu\sigma\rho}R^{\mu\nu\sigma\rho}+R^2-4R_{\mu\nu}R^{\mu\nu})-4\pi G F_{\mu\nu}F^{\mu\nu}\bigg].
\end{equation}
The equation incorporates the electromagnetic field tensor, represented as $F_{\mu\nu} = \partial_{\mu}A_{\nu} - \partial_{\nu}A_{\mu}$, alongside the Gauss-Bonnet coupling constant $\alpha_{GB}$, Newton's constant $G$, and the cosmological constant $\Lambda$. In string theory, the Gauss-Bonnet coupling constant is positive with dimensions of $(length)^2$. Here, we focus on a spherical horizon topology, which corresponds to $k = 1$. As a result, the metric can be expressed in the following form,
\begin{equation}\label{eq2}
ds^2=-f(r)dt^2+f^{-1}(r) dr^2+r^2(d\theta^2+\sin^2\theta d\phi^2+\cos^2\theta d\Omega^2_{d-4}),
\end{equation}
and,
 \begin{equation}\label{eq3}
f(r)=1+\frac{r^2}{2\beta_0}\big[1-\bigg(1+\frac{64\pi \beta_0 GM}{(d-2)\Sigma r^{d-1}}-\frac{8G\beta_0 Q^2 }{(d-2)(d-3)r^{2(d-2)}}-\frac{4\beta_0}{\ell^2}\bigg)^{\frac{1}{2}}\big],
\end{equation}
The quantities $\Sigma$, $Q$, and $M$ denote the area of the $d-2$ dimentional unit sphere, the electrice charhe, and the black hole's mass, respectively. Furthermore, $\beta_0=(d-4)(d-3) \alpha_{GB}$.
By solving the equation $f(r_+)=0$ and using the relation $T=\frac{f^{\prime}(r_+)}{4\pi}$, the mass and temperature of the black hole can calculated \cite{24,25,26,27}, 
\begin{equation}\label{eq4}
M=\frac{\Sigma (d-2) r_+^{d-3}}{16\pi G r_+^2 \ell^2}(r_+^2 \ell^2+\beta \ell^2 + r_+^4)+\frac{\Sigma Q^2}{8\pi (d-3)r_+^{d-3}},
\end{equation}
and,
 \begin{equation}\label{eq5}
T=\frac{(d-1)(d-2)r_+^{2(d-2)}+(d-2)(d-3)\ell^2 r_+^{2(d-3)}+(d-5)(d-2)\beta_0 r_+^{2(d-4)}-2GQ^2}{4\pi r_+(r_+^2+2\beta_0)\ell^2(d-2)r_+^{2(d-4)}},
\end{equation}
The entropy of the black hole can also be represented as cite{26},
 \begin{equation}\label{eq6}
S=\frac{\Sigma r_+^{d-2}}{4G}\bigg(1+\frac{2(d-2)\beta_0}{(d-4)r_+^2} \bigg).
\end{equation}
We then analyze the CFT thermodynamic properties of a five-dimensional $(d=5)$ Gauss-Bonnet black hole.
\subsection{CFT thermodynamics for GB-AdS black hole with $d=5$}
To study the thermodynamics of a charged AdS black hole within the context of CFT, we employ holographic correspondences that link bulk and boundary quantities. In this analysis, the curvature radius of the boundary, denoted as $R$, is distinct from the AdS radius $\ell$ in the bulk. The CFT metric, defined by its conformal scaling invariance, is expressed accordingly\cite{28,29,30},
\begin{equation}\label{eq7}
\begin{split}
ds^2=\omega^2(-dt^2+\ell^2 d\Omega_{d-2}^2).
\end{split}
\end{equation}
The dimensionless conformal factor, denoted as $\omega$, is allowed to vary freely, reflecting the conformal symmetry inherent in the boundary theory. In the spherical case, $d\Omega_{d-2}^2$ represents the metric of a $(d-2)$-dimensional sphere, with its volume denoted by $\Omega_{d-2}=k$. It is assumed that $\omega$ does not depend on the boundary coordinates. Under these assumptions, the volume of the conformal field theory (CFT) is determined accordingly\cite{31,32},
\begin{equation}\label{eq8}
\begin{split}
\mathcal{V}=k R^{d-2},
\end{split}
\end{equation}
The variable curvature radius of the manifold hosting the CFT is denoted as $R = \omega \ell$. Consequently, the variation in the CFT volume $\mathcal{V}$ remains entirely unaffected by changes in the central charge $C$. This central charge exhibits a dual relationship in the framework of Einstein's gravity, as demonstrated in the following equation \cite{31,32,33},
\begin{equation}\label{eq999}
C=\frac{k \ell^{d-2}}{16 \pi G}.
\end{equation}
We keep Newton's constant $G$ fixed while allowing the bulk curvature radius, $ \ell$, to vary, thereby altering the central charge $C$ as outlined in equation \eqref{eq999}.
The first law of thermodynamic in the CFT case for a five-dimensional GB-AdS black hole and the correspondence between the bulk and boundary values are obtained as follows \cite{33},
\begin{equation}\label{eq10}
\begin{split}
  \delta E= \tilde{T} \delta \tilde{S}+\mu \delta C-p \delta \mathcal{V}+ \tilde{\Phi} \delta \tilde{Q}+ \tilde{A} \delta\tilde{\beta_0}
\end{split}
\end{equation}
\begin{equation}\label{eq1111}
\begin{split}
 E=\frac{M}{\omega}, \qquad  \tilde{S}=S,\qquad \tilde{T}=\frac{T}{\omega}, \qquad  \tilde{\Phi}=\frac{\Phi\sqrt{G}}{\omega\ell},\qquad \tilde{Q}=\frac{Q\ell}{\sqrt{G}}, \qquad \tilde{\mathcal{A}}=\frac{\mathcal{A}}{\omega\ell},\qquad \tilde{\beta_0}=\ell\beta_0,
\end{split}
\end{equation}
\begin{equation}\label{eq12}
\begin{split}
&  \mu=\frac{1}{C}(E-\tilde{T}S-\tilde{\Phi}\tilde{Q}-\tilde{\mathcal{A}}\tilde{\beta_0})\\
&p=\frac{E}{3\mathcal{V}} , \qquad \mathcal{V}=k R^3
\end{split}
\end{equation}
Where $\mu$ is the thermodynamic conjugate of the central charge, i.e., the chemical potential, $\Phi$ is the electric potential, and $\mathcal{A}$ is the thermodynamic conjugate of the Gauss-Bonnet parameter. Also, equation \eqref{eq1111} establishes the correspondence between bulk and boundary quantities.
We aim to derive the internal energy formula for the GB black hole in five dimensions within the framework of CFT.
For $d = 5$, we find that $\beta_0 = 2\alpha$. To determine the internal energy formula for the CFT, we define two dimensionless parameters as follows\cite{33},
\begin{equation}\label{eq13}
\begin{split}
x\equiv \frac{r_+}{\ell} \qquad \qquad  y\equiv \frac{\tilde{\beta_0}}{\ell^3},
\end{split}
\end{equation}
Thus, based on equations \eqref{eq4}, \eqref{eq6}, \eqref{eq999}, \eqref{eq1111}, and \eqref{eq13}, we obtain,
\begin{equation}\label{eq14}
\begin{split}
&E=\frac{\Sigma  \left(k^2 \tilde{Q}^2+768 \pi ^2 C^2 x^6+768 \pi ^2 C^2 x^4+768 \pi ^2 C^2 x^2 y\right)}{256 \pi ^2 C (k^{2} \mathcal{V})^{\frac{1}{3}} x^2}\\
&\tilde{S}=\frac{4 \pi  C \Sigma  \left(x^3+6 x y\right)}{k}.
\end{split}
\end{equation}
Additionally, we can determine the thermodynamic properties of the CFT,
\begin{equation}\label{eq15}
\begin{split}
&\tilde{T}=\bigg(\frac{\partial E}{ \partial \tilde{S}} \bigg)_{\tilde{Q},\mathcal{V},C,\tilde{\beta}}= \bigg(\frac{k}{\mathcal{V}}\bigg)^{\frac{1}{3}}\bigg(\frac{-k^2 \tilde{Q}^2+1536 \pi ^2 C^2 x^6+768 \pi ^2 C^2 x^4}{1536 \pi ^3 C^2 x^3 \left(x^2+2 y\right)}\bigg),
\end{split}
\end{equation}
\begin{equation}\label{eq16}
\begin{split}
\tilde{\Phi}=\bigg(\frac{\partial E}{ \partial \tilde{Q}} \bigg)_{\tilde{S},\mathcal{V},C,\tilde{\beta}}=\bigg(\frac{k^4}{\mathcal{V}}\bigg)^{\frac{1}{3}}\bigg(\frac{\Sigma  \tilde{Q}}{128 \pi ^2 c x^2} \bigg),
\end{split}
\end{equation}
\begin{equation}\label{eq17}
\begin{split}
p=-\bigg(\frac{\partial E}{ \partial \mathcal{V}} \bigg)_{\tilde{S},\tilde{Q},C,\tilde{\beta}}=\frac{E}{2\mathcal{V}},
\end{split}
\end{equation}
\begin{equation}\label{eq18}
\begin{split}
\mu=\bigg(\frac{\partial E}{ \partial C} \bigg)_{\tilde{S},\tilde{Q},\mathcal{V},\tilde{\beta}}=\frac{\Sigma  \left(-k^2 \tilde{Q}^2+768 \pi ^2 C^2 x^6+768 \pi ^2 C^2 x^4+768 \pi ^2 C^2 x^2 y\right)}{256 \pi ^2 C^2 k^{2/3} \sqrt[3]{\mathcal{V}} x^2},
\end{split}
\end{equation}
\begin{equation}\label{eq19}
\begin{split}
\tilde{\mathcal{A}}=\bigg(\frac{\partial E}{ \partial \tilde{\beta}} \bigg)_{\tilde{S},\tilde{Q},\mathcal{V},C}=\frac{3 C \Sigma  y}{\tilde{\beta} \sqrt[3]{k^{2}\mathcal{V}}}.
\end{split}
\end{equation}
Here, equation \eqref{eq17} represent the equation of state in CFT.
The concept of a variable Newton's constant is relevant exclusively within the framework of the mixed first law, which modifies the volume term in the black hole phase transition formula. Conversely, the internal energy formula, the boundary first law, and the Euler relation do not require a dynamic Newton's constant. The primary objective is to establish a precise dual correspondence between the bulk and boundary first laws. To resolve the degeneracy between the volume ($\mathcal{V}$) and the central charge ($C$), a dynamic parameter, $\omega$, is introduced. These variables are employed to examine the various stages across different thermodynamic ensembles within the dual CFT framework.
In CFT thermodynamics, the phase structure of a black hole can be analyzed through the Helmholtz free energy, defined as:
\begin{equation}\label{eq20}
\begin{split}
\mathcal{F}=E-\tilde{T}\tilde{S}.
\end{split}
\end{equation}
\section{Thermodynamic topology}
Recent advancements have introduced innovative methods for analyzing and identifying critical points and phase transitions in black hole thermodynamics. One notable approach is the topological method, which employs Duan’s topological current $\phi$-mapping theory to offer a topological perspective on thermodynamics\cite{10,11}. To examine the thermodynamic properties of black holes, various quantities such as mass and temperature are utilized to describe the generalized free energy. Given the relationship between mass and energy in black holes, the generalized free energy function is expressed as a standard thermodynamic function. The Euclidean time period $\tau$ and its inverse, the temperature $T$, are essential components in this formulation. The generalized free energy is considered on-shell only when $\tau$ equals the inverse of the Hawking temperature\cite{10,11}. A vector $\phi$ is constructed to facilitate this analysis, with components derived from the partial derivatives of the generalized free energy. The direction of this vector is significant, as it points outward at specific angular positions, indicating the ranges for the horizon radius and angular coordinates. Using Duan's $\phi$-mapping topological current theory, a topological current can be defined, which is conserved according to Noether's theorem. To determine the topological number, the topological current is reformulated, incorporating the Jacobi tensor. This tensor simplifies to the standard Jacobi form under certain conditions, and the conservation equation reveals that the topological current is non-zero only at specific points. Through detailed calculations, the topological number or total charge $W$ can be expressed, involving the Hopf index and the sign of the topological current at zero points. The winding number, which is independent of the region's shape, directly relates to black hole stability. A positive winding number corresponds to a stable black hole state, while a negative winding number indicates instability. This topological approach provides a robust framework for understanding the stability and phase transitions of black holes, offering new insights into their thermodynamic behavior. The generalized free energy is determined as follows\cite{10,11}:
\begin{equation}\label{F1}
\mathcal{F} = M - \frac{S}{\tau},
\end{equation}
In this context, $\tau$ represents the Euclidean time period, and its inverse, $T$, represents the system's temperature. The generalized free energy is considered on-shell only when $\tau$ matches the inverse of the Hawking temperature.
By applying CFT thermodynamics to examine the phase structure of black holes, we redefine the generalized Helmholtz energy within the CFT framework as follows:
\begin{equation}\label{F111}
\begin{split}
\mathcal{F} = E - \frac{\tilde{S}}{\tilde{\tau}}.
\end{split}
\end{equation}
In this context, the generalized Helmholtz energy is on-shell only when $(\tilde{\tau} = \tilde{\tau}_{H} = \frac{1}{\tilde{T}_{H}})$.
To facilitate this analysis, a vector $(\phi)$ is constructed with components derived from the partial derivatives as follows:
\begin{equation}\label{F2}
\phi = \left(\frac{\partial \mathcal{F}}{\partial r_{H}}, -\cot \Theta \csc \Theta \right).
\end{equation}
In this scenario, $(\phi^{\Theta})$ becomes infinite, and the vector points outward at the angles $(\Theta = 0)$ and $(\Theta = \pi)$. The permissible ranges for the horizon radius $(r_{H})$ and the angle $(\Theta)$ are from 0 to infinity and from 0 to $(\pi)$, respectively. By applying Duan's $(\phi)$-mapping topological current theory, we can define a topological current as follows:
\begin{equation}\label{F3}
j^{\mu} = \frac{1}{2\pi} \varepsilon^{\mu\nu\rho} \varepsilon_{ab} \partial_{\nu} n^{a} \partial_{\rho} n^{b}, \quad \mu, \nu, \rho = 0, 1, 2,
\end{equation}
In this formulation, $n$ is defined as $(n^1, n^2)$, where $(n^1 = \frac{\phi^r}{|\phi|})$ and $(n^2 = \frac{\phi^\Theta}{|\phi|})$. According to the conservation equation, the current $(j^{\mu})$ is non-zero exclusively at the points where $(\phi = 0)$. After performing the necessary calculations, the topological number or total charge $W$ can be determined as follows:
\begin{equation}\label{F4}
W = \int_{\Sigma} j^{0} d^2 x = \sum_{i=1}^{n} \beta_{i} \eta_{i} = \sum_{i=1}^{n} \omega_{i}.
\end{equation}
In this context, $(\beta_i)$ represents the positive Hopf index, which counts the number of loops made by the vector $(\phi^a)$ in the $(\phi)$-space when $(x^\mu)$ is close to the zero point $(z_i)$. Meanwhile, $(\eta_i)$ is defined as the sign of $(j^0(\phi/x)_{z_i})$, which can be either +1 or -1. The term $(\omega_i)$ denotes the winding number associated with the $i$-th zero point of $(\phi)$ within the region $(\Sigma)$.
\section{Non-extensive Entropy}
Non-extensive entropy, proposed by Tsallis, is an extension of the traditional Boltzmann-Gibbs entropy. This concept is particularly valuable for systems with non-linear dynamics and strong initial condition dependencies. Unlike Boltzmann-Gibbs entropy, which assumes linear scaling with system size, non-extensive entropy can manage systems where this linearity is absent. This makes it relevant to various fields, including theoretical physics, cosmology, and statistical mechanics, especially for systems with long-range interactions, fractal structures, or memory effects\cite{a33,a33',a33''}.\\\\
\quad$\bullet$ R\'{e}nyi entropy is a type of non-extensive entropy applied in black hole thermodynamics. It is characterized by a parameter that adjusts the degree of non-extensiveness, which must be within a specific range to ensure the entropy function is well-defined. When used for black holes, R\'{e}nyi entropy offers a framework for understanding their thermodynamic properties, extending traditional Boltzmann-Gibbs statistics\cite{a25,a26,a27}.
\begin{equation}\label{N1}
\begin{split}
S_R = \frac{1}{\lambda} \ln(1 + \lambda S_{BH})
\end{split}
\end{equation}
The parameter $(\lambda)$ in non-extensive entropy is essential for defining the entropy function. For the entropy function to be well-defined, $(\lambda)$ must be within the range $(-\infty < \lambda < 1)$. Values outside this range make the entropy function convex and thus ill-defined. In black hole thermodynamics using R\'{e}nyi statistics, the entropy $(S_R)$ is properly defined when $(\lambda)$ is between 0 and 1. Within this range, $(\lambda)$ shows favorable thermodynamic properties, as recent studies have demonstrated. Notably, as the R\'{e}nyi parameter $(\lambda)$ approaches zero, the generalized off-shell free energy converges to classical Boltzmann-Gibbs statistics.\\\\
\quad$\bullet$ Sharma-Mittal entropy is another significant form of non-extensive entropy, generalizing both R\'{e}nyi and Tsallis entropies. This entropy has been particularly useful in cosmological studies, such as explaining the accelerated expansion of the universe by effectively utilizing vacuum energy. Although non-extensive entropies have been used to study black holes, Sharma-Mittal entropy has not been extensively applied in this context. This presents an opportunity to explore the thermodynamic properties of black holes using Sharma-Mittal entropy, considering them as strongly coupled gravitational systems\cite{a28,a29,a30}.
\begin{equation}\label{N2}
\begin{split}
S_{SM} = \frac{1}{\alpha} \left( (1 + \beta S_T)^\frac{\alpha}{\beta} - 1 \right).
\end{split}
\end{equation}
In this context, $S_T$ represents the Tsallis entropy, derived from the horizon area $(A = 4\pi r^2)$, where $r$ is the radius of the black hole's event horizon. The parameters $\alpha$ and $(\beta)$ are adjustable and need to be calibrated using observational data. Interestingly, when $\alpha$ approaches zero, Sharma-Mittal entropy simplifies to R\'{e}nyi entropy. Similarly, when $\alpha$ equals $(\beta)$, it reduces to Tsallis entropy.\\\\
\quad$\bullet$ Non-extensive statistical mechanics in Loop Quantum Gravity provides the entropy\cite{a40,a50,a60}
\begin{equation}\label{N3}
S_q = \frac{1}{1 - q} \left[ e^{(1-q)\Lambda(\gamma_0)S} - 1 \right],
\end{equation}
where the entropic index $q$ quantifies how the probability of frequent events is enhanced relative to infrequent ones,
\begin{equation}\label{N4}
\Lambda(\gamma_0) = \frac{\ln 2}{\sqrt{3} \pi \gamma_0},
\end{equation}
and $\gamma_0$ is the Barbero-Immirzi parameter, typically assumed to take one of the two values $\frac{\ln 2}{\pi \sqrt{3}}$ or $\frac{\ln 3}{2\pi \sqrt{2}}$, depending on the gauge group used in Loop Quantum Gravity. However, $\gamma_0$ is a free parameter in scale-invariant gravity\cite{a70,a80,a90,a100}. With the first choice of $\gamma_0$, $\Lambda(\gamma_0)$ becomes unity, and the entropy reduces to the Bekenstein-Hawking form in the limit $q \to 1$, corresponding to extensive statistical mechanics. This Loop Quantum Gravity entropy has been applied to black holes in\cite{a40,a50} and to cosmology in\cite{a60}.
\section{Thermodynamic topology and the R\'{e}nyi entropy}
The application of R\'{e}nyi entropy in black hole models can yield interesting consequences and provide new insights into the complex structure of general relativity. Based Eqs. \ref{eq14}, \ref{F111} and \ref{N1}, the generalized Helmholtz free energy we have:
\begin{equation}\label{R1}
\mathcal{F }=\frac{\left(768 x^{6} C^{2} \pi^{2}+768 x^{4} C^{2} \pi^{2}+768 y \,x^{2} C^{2} \pi^{2}+Q^{2} k^{2}\right) \Sigma}{256 x^{2} \left(\mathcal{V} \,k^{2}\right)^{\frac{1}{3}} C \pi^{2}}-\frac{\ln \! \left(1+\frac{4 \left(x^{3}+6 y x \right) \Sigma  C \pi  \lambda}{k}\right)}{\lambda  \tilde{\tau}}
\end{equation}
With respect to Bekenstein-Hawking entropy,
\begin{equation}\label{R2}
S_{\mathit{BH}}=\frac{4 \left(x^{3}+6 y x \right) \Sigma  C \pi}{k}.
\end{equation}
Also, the R\'{e}nyi entropy will be,
\begin{equation}\label{R3}
S_{R}=\frac{\ln \! \left(1+\frac{4 \left(x^{3}+6 y x \right) \Sigma  C \pi  \lambda}{k}\right)}{\lambda}.
\end{equation}
Now form Eq.\ref{F2} for components of the vector field $\Phi$ and $\tilde{\tau}$ we have:
\begin{equation*}\label{(0)}
\varphi_{1}=x^{3} C^{2} \left(-4 x^{2} \tilde{\tau}  \left(x^{2}+\frac{1}{2}\right) \lambda  \left(x^{2}+6 y \right) \Sigma  C +\left(x^{2}+2 y \right) \left(\mathcal{V} \,k^{2}\right)^{\frac{1}{3}}\right) \pi^{3}-x^{4} k \,C^{2} \left(x^{2}+\frac{1}{2}\right) \tilde{\tau}  \,\pi^{2},
\end{equation*}
\begin{equation}\label{(4)}
\phi^{x}=-\frac{3 \left(\varphi_{1}+\frac{C \,Q^{2} \Sigma  k^{2} \lambda  \tilde{\tau}  x \left(x^{2}+6 y \right) \pi}{384}+\frac{Q^{2} k^{3} \tilde{\tau}}{1536}\right) \Sigma}{\left(\mathcal{V} \,k^{2}\right)^{\frac{1}{3}} \left(C \lambda  x \Sigma  \left(x^{2}+6 y \right) \pi +\frac{k}{4}\right) x^{3} \tilde{\tau}  C \pi^{2}},
\end{equation}
\begin{equation}\label{(5)}
\phi^{\theta}=-\frac{\cos \! \left(\theta \right)}{\sin \! \left(\theta \right)^{2}}
\end{equation}
\begin{equation*}\label{(0)}
\tilde{\tau_{1}}=6144 C^{3} \pi^{3} \Sigma  \lambda  x^{9}+36864 C^{3} \pi^{3} \Sigma  \lambda  x^{7} y +3072 C^{3} \pi^{3} \Sigma  \lambda  x^{7}+18432 C^{3} \pi^{3} \Sigma  \lambda  x^{5} y +1536 C^{2} \pi^{2} k \,x^{6}
\end{equation*}
\begin{equation}\label{(6)}
\tilde{\tau}=\frac{1536 \left(x^{2}+2 y \right) C^{2} \pi^{3} x^{3} \left(\mathcal{V} \,k^{2}\right)^{\frac{1}{3}}}{-4 C \pi  Q^{2} \Sigma  k^{2} \lambda  x^{3}-24 C \pi  Q^{2} \Sigma  k^{2} \lambda  x y +768 C^{2} \pi^{2} k \,x^{4}-Q^{2} k^{3}+\tilde{\tau}_{1}}
\end{equation}
In order to better observe the effect of the R\'{e}nyi entropy and it's parameter changes on phase transition, we divide our studies into two parts. In the first case, we will examine the phase
behavior of the sample below critical $C$, and in the second case, we will go to values greater than critical $C$.
\subsection{$ C=0.15 <C_{c} $}
First, it is better to examine the effect of changing $\lambda$ on $\tilde{\tau}$ (the temperature inverse).
\begin{figure}[H]
 \begin{center}
 \subfigure[]{
 \includegraphics[height=6.5cm,width=8cm]{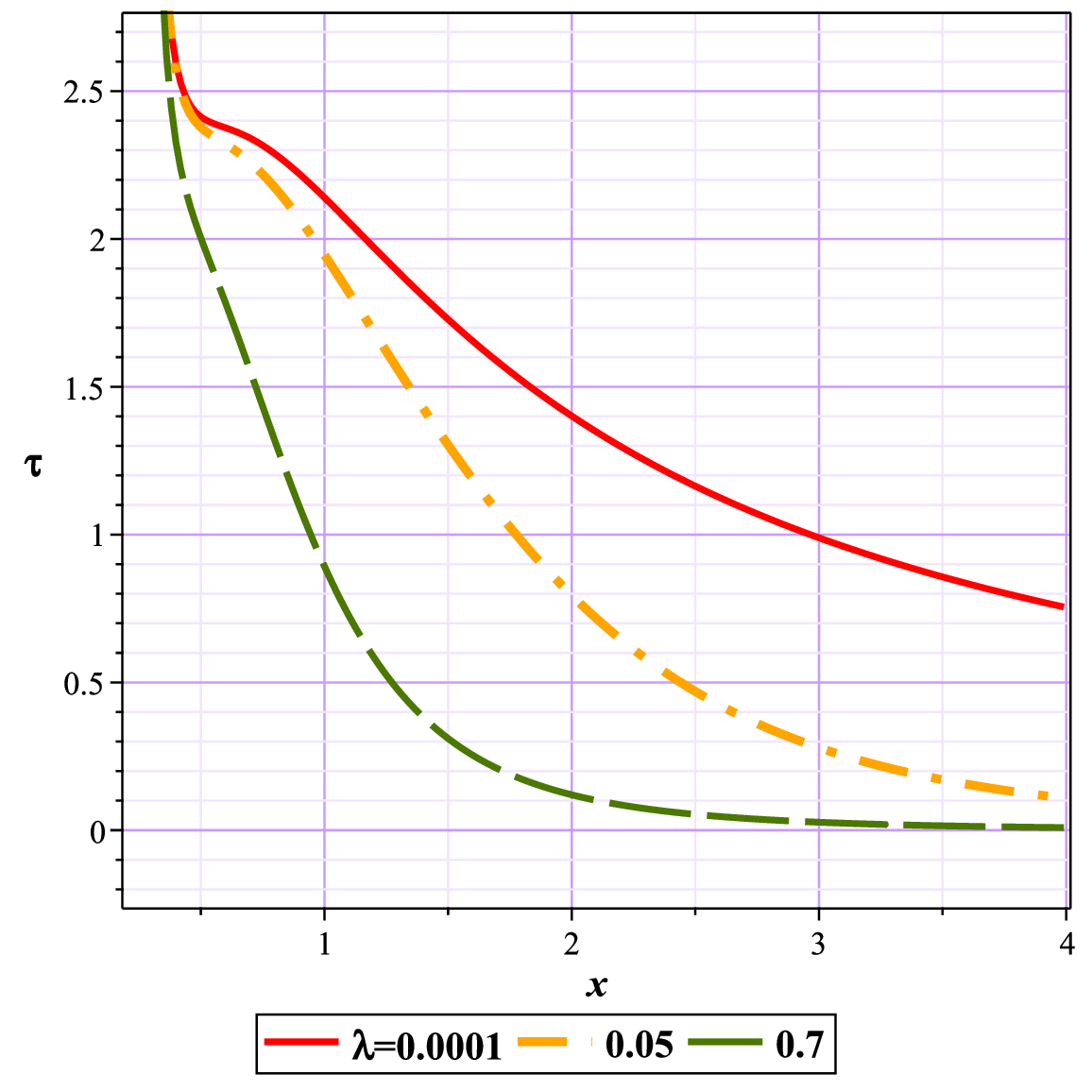}
 \label{400a}}
 \caption{\small{$(\tilde{\tau}$ VS x) with different $\lambda$ for 5D G-B black hole model and R\'{e}nyi entropy }}
 \label{m400}
\end{center}
\end{figure}
As can be seen in Fig. \ref{m400}, in this case, changes in $\lambda$ do not seem to cause the $\tau$ function to have an extremum. Therefore, we expect that for any choice of value for this function, a topological charge of +1 will always appear, which can be clearly seen in Fig. \ref{m401}. In fact, below the critical C, the sample will seem to experience only one second-order phase transition.
\begin{figure}[H]
 \begin{center}
 \subfigure[]{
 \includegraphics[height=6.5cm,width=8cm]{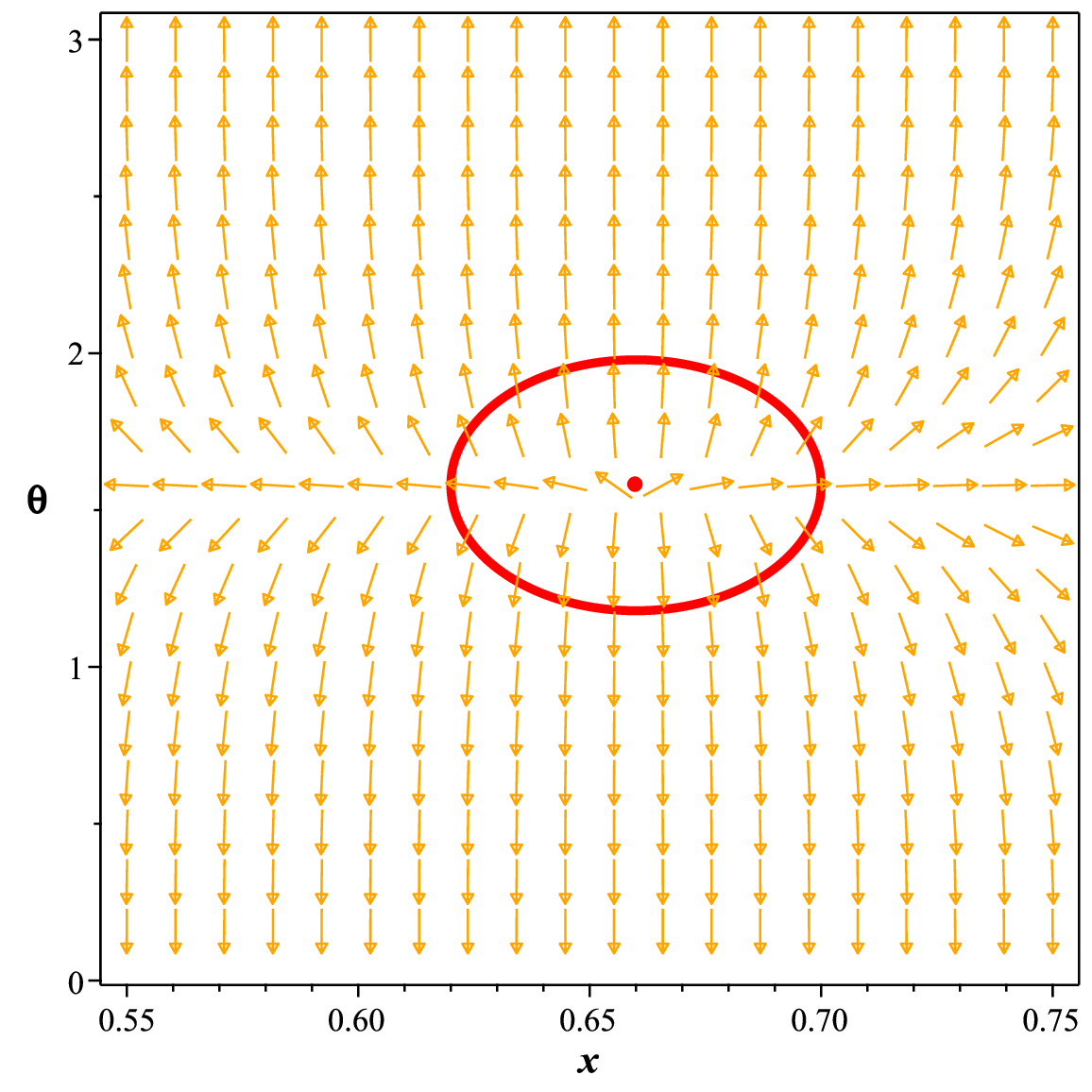}
 \label{401}}
 \caption{\small{The normal vector field $n$ in the $(x-\theta)$ plane. The ZP(Zero Points) is located at $ (x,\theta)=(0.66,1.57)$ with respect to $(\lambda = 0.0001, \mathcal{V}= 1, k = 1, Q = 1, \Sigma = 1, C = 0.15, y = 0.01 )$  }}
 \label{m401}
\end{center}
\end{figure}
\subsection{$ C=2.5 >C_{c} $}
At values greater than the critical $C$, the black hole's phase behavior appears to be completely different.
\begin{figure}[H]
 \begin{center}
 \subfigure[]{
 \includegraphics[height=6.5cm,width=8cm]{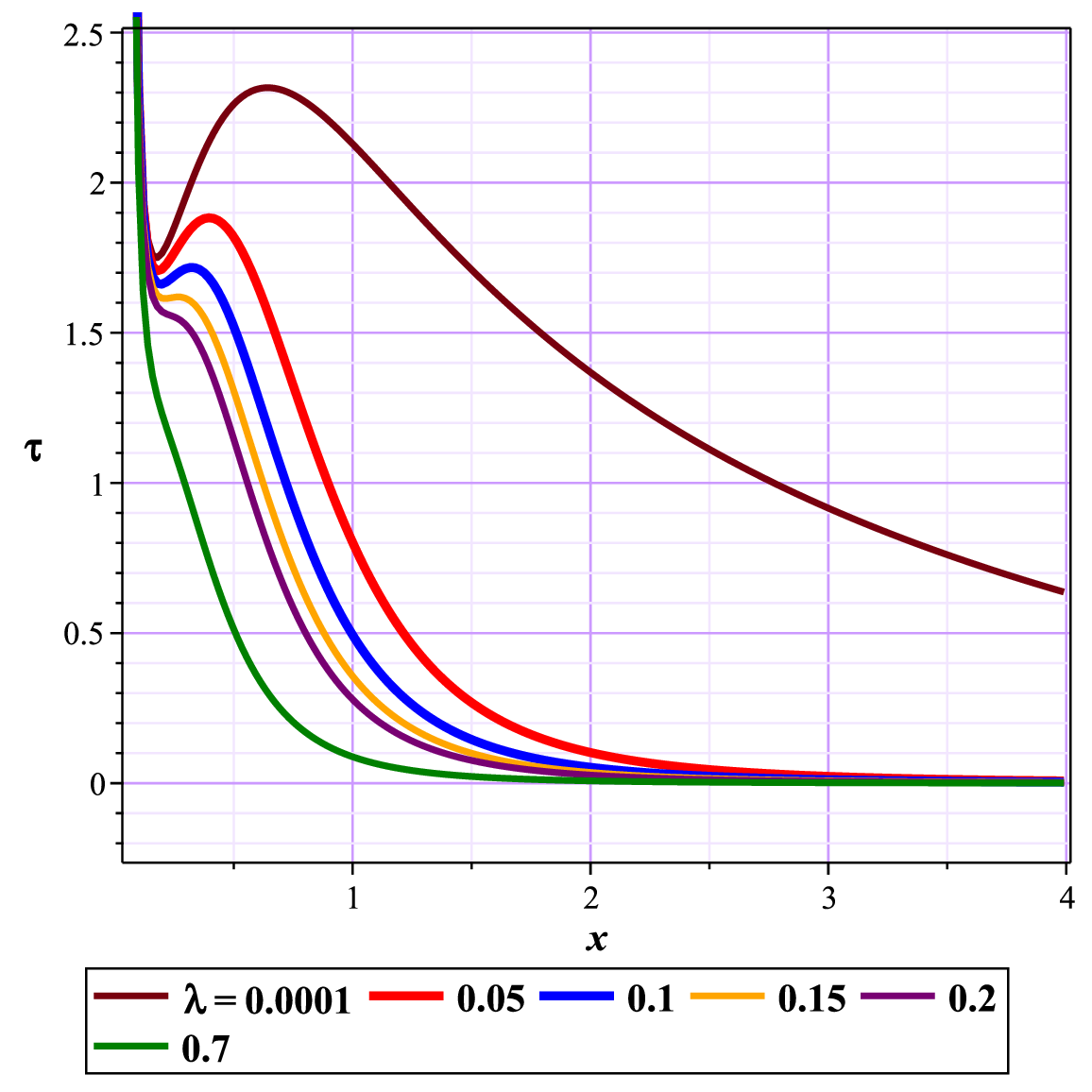}
 \label{403}}
 \caption{\small{($\tau$ VS x) with different $\lambda$  for 5D G-B black hole model and R\'{e}nyi entropy }}
 \label{m403}
\end{center}
\end{figure}
As can be seen in Fig. \ref{m403}, the phase behavior of the black hole will be affected by the choice of the $\lambda$ parameter and clearly, depending on our choice, each can experience both forms of phase transition. At very small $\lambda$ (burgundy line), the intensity of the changes is very high. We know that this very small $\lambda$ means moving towards the Bekenstein-Hawking entropy. In other words, it can be said with certainty that the structure in the usual Bekenstein-Hawking entropy and below the critical C will definitely experience a first-order phase transition.
But in the case of our study, for example for $\lambda = 0.05 $ (red line), we can clearly see that extrema in the $\tilde{\tau}$ function have appeared. This means that a ZP with topological charge -1(green contoure) will definitely appear in the topological study, Fig. \ref{m404}.
This means that our black hole will pass through an unstable intermediate black hole (which appears and then disappears) during the phase transition from a small to a large black hole, and will experience a creation point and an annihilation point in this transition.
\begin{figure}[H]
 \begin{center}
 \subfigure[]{
 \includegraphics[height=6.5cm,width=8cm]{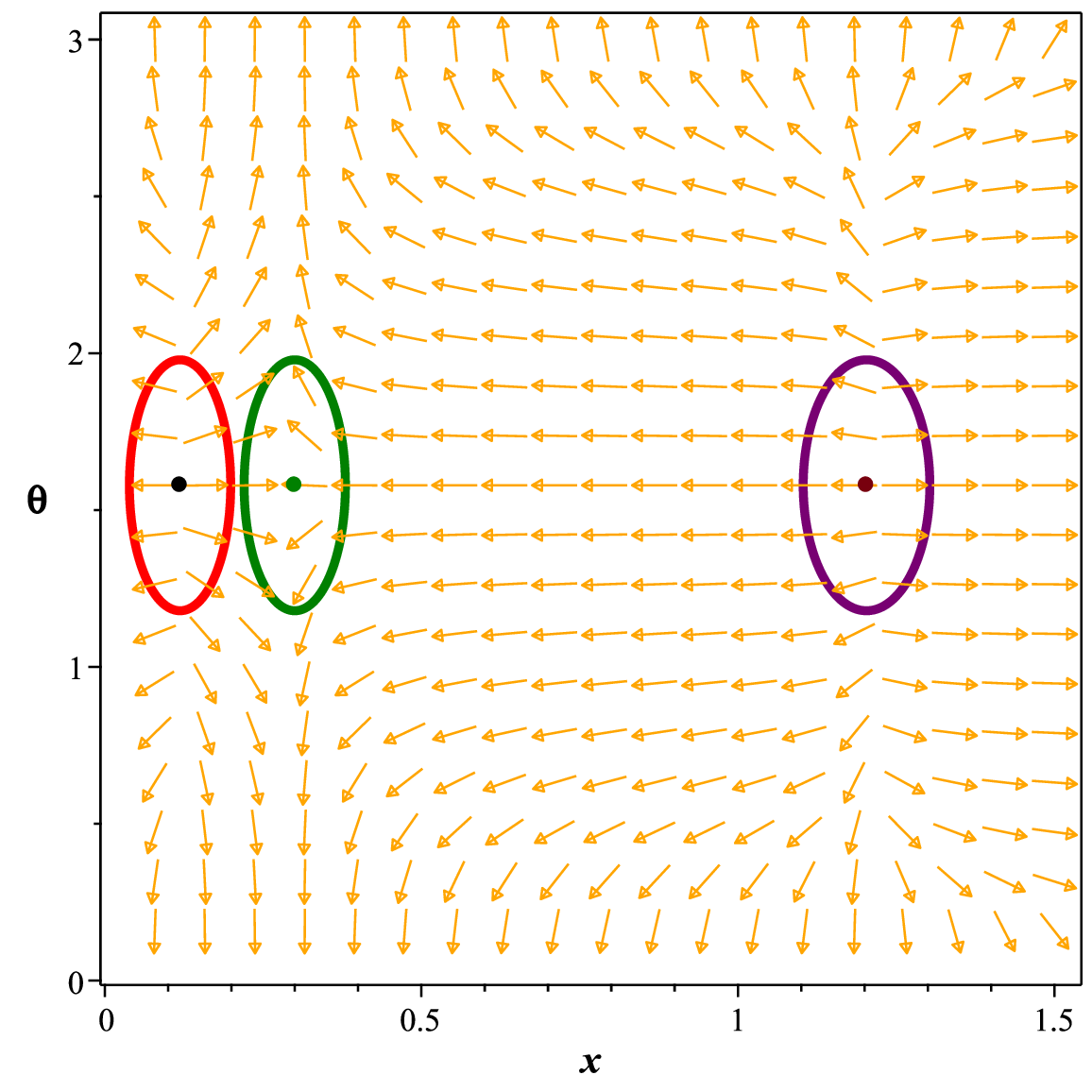}
 \label{404}}
 \caption{\small{The normal vector field $n$ in the $(x-\theta)$ plane. The ZPs are located at $ (x,\theta)=(0.1187834875,1.57)$ , $ (x,\theta)=(0.2999999998,1.57)$, $ (x,\theta)=(1.203156012,1.57)$  with respect to $(\lambda = 0.0001, \mathcal{V}= 1, k = 1, Q = 1, \Sigma = 1, C = 2.5, y = 0.01, \tilde{\tau} = 1.956446998 )$  }}
 \label{m404}
\end{center}
\end{figure}
As can be seen in Fig. \ref{m403}, around $\lambda=0.15$, the extrema of the $\tilde{\tau}$ function disappear and the intermediate black hole is effectively eliminated. This means that the phase transition takes on a second-order form and a continuous gradual transformation will transform the small black hole into a large black hole. This trend is also true for values larger than $0.15$. 
\section{Thermodynamic topology and Sharma-Mittal entropy}
Using equations \eqref{eq14}, \eqref{eq15}, and \eqref{F111}, we can rewrite the generalized Helmholtz energy for the Sharma-Mittal entropy as follows,
\begin{equation}\label{eq691}
\begin{split}
\mathcal{F}=\frac{\Sigma  \left(768 C^2 \left(x^4+x^2+y\right)+\frac{k^2 Q^2}{\pi ^2 x^2}\right)}{256 C k^{2/3} \sqrt[3]{\mathcal{V}}}- \frac{\left(\frac{4 \pi  \beta  C \Sigma  \left(x^3+6 x y\right)}{k}+1\right)^{\alpha /\beta }-1}{\alpha  \tilde{\tau} }
\end{split}
\end{equation}
From \eqref{F2}, for the components of the vector field $\Phi$ and $\tilde{\tau}$, we obtain:
\begin{equation}\label{eq692}
\begin{split}
\phi ^{x}=\frac{6 C \Sigma }{\ell}\bigg(\frac{2 x^3+x}{k^{2/3} \sqrt[3]{\mathcal{V}}}-\frac{2 \pi  \left(x^2+2 y\right) \left(\frac{4 \pi  \beta  C \Sigma  x \left(x^2+6 y\right)}{k}+1\right)^{\frac{\alpha }{\beta }-1}}{k \tilde{\tau} }\bigg)
\end{split}
\end{equation}
\begin{equation}\label{eq693}
\begin{split}
\phi ^{\theta}=-\frac{\cot (\theta )}{\sin (\theta )}
\end{split}
\end{equation}
\begin{equation}\label{eq694}
\begin{split}
\tilde{\tau} =\frac{1536 \pi ^3 C^2 \sqrt[3]{\mathcal{V}} x^3 \left(x^2+2 y\right) \left(\frac{4 \pi  \beta  C \Sigma  x \left(x^2+6 y\right)}{k}+1\right)^{\frac{\alpha }{\beta }-1}}{\sqrt[3]{k} \left(768 \pi ^2 C^2 x^4 \left(2 x^2+1\right)-k^2 Q^2\right)}
\end{split}
\end{equation}
\subsection{$C=0.15 <C_{c}$}
\begin{figure}[h!]
 \begin{center}
 \subfigure[]{
 \includegraphics[height=6.5cm,width=8cm]{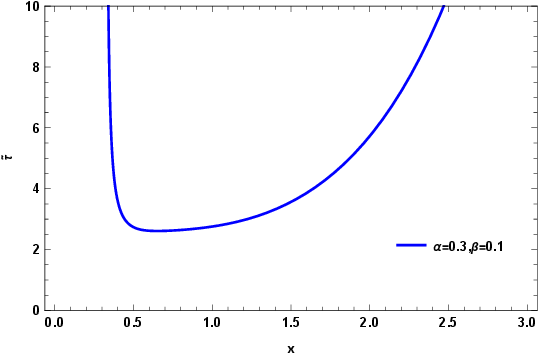}
 \label{fig101a}}
 \subfigure[]{
 \includegraphics[height=6.5cm,width=8cm]{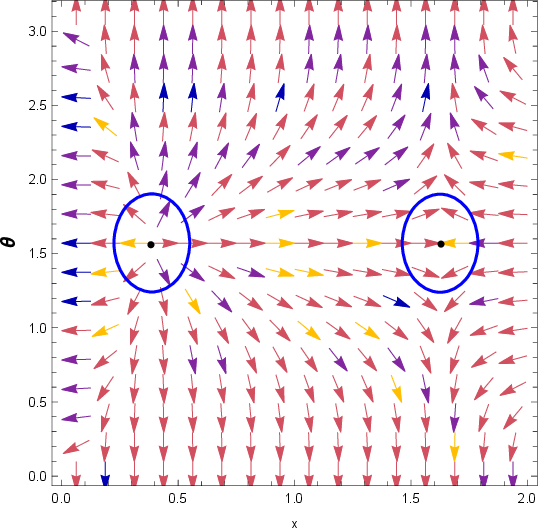}
 \label{fig101b}}
 \caption{\small{ (a)  The plot of the curve of equation \eqref{eq694} with respect to $k=\mathcal{V}=Q=\ell=\Sigma=1$, $y=0.01$, $\alpha=0.3, \beta=0.1$ and $C=0.1<C_c=0.187$ (b) , the blue arrows represent the vector field n on a portion of the  $(x - \theta)$  plane for the quantum-corrected (AdS-RN) black holes in Kiselev spacetime. The blue loops enclose the ZPs $\tilde{\tau}=4$.}}
 \label{fig100}
 \end{center}
 \end{figure}

\begin{figure}[h!]
 \begin{center}
 \subfigure[]{
 \includegraphics[height=4cm,width=5cm]{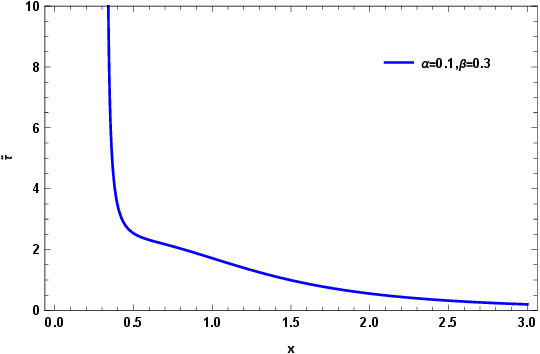}
 \label{fig201a}}
 \subfigure[]{
 \includegraphics[height=4cm,width=5cm]{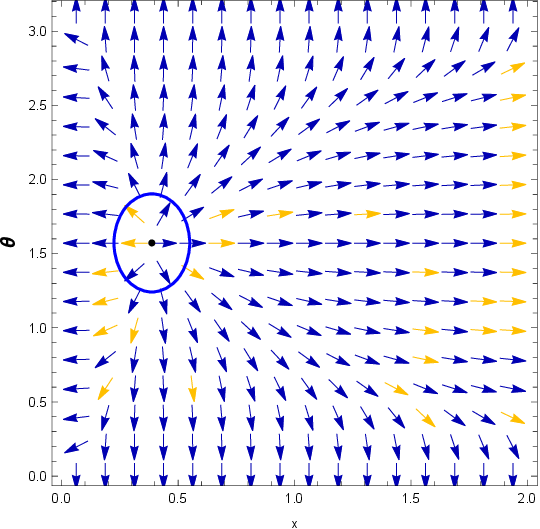}
 \label{fig201b}}
 \caption{\small{ (a)  The plot of the curve of equation (17) with respect to $k=\mathcal{V}=Q=\ell=\Sigma=1$, $y=0.01$, $\alpha=0.1, \beta=0.3$ and $C=0.1<C_c=0.187$ (b), the blue arrows represent the vector field n on a portion of the $( x - \theta)$ plane for the Einstein-Gauss-Bonnet black holes. The blue loops enclose the ZPs $\tilde{\tau}=4$.}}
 \label{fig200}
 \end{center}
 \end{figure}

In this scenario, where no first-order phase transition takes place, two distinct topological charges can exist: for $\alpha>\beta$ (Fig. \eqref{fig100}), the total topological charge is +1, while for $\alpha\leq\beta$ (Fig. \eqref{fig200}), it is 0.
Next, we will analyze the scenario in which a first-order phase transition takes place.
\subsection{$C=2.5 >C_{c}$}
\begin{figure}[h!]
 \begin{center}
 \includegraphics[height=6cm,width=7cm]{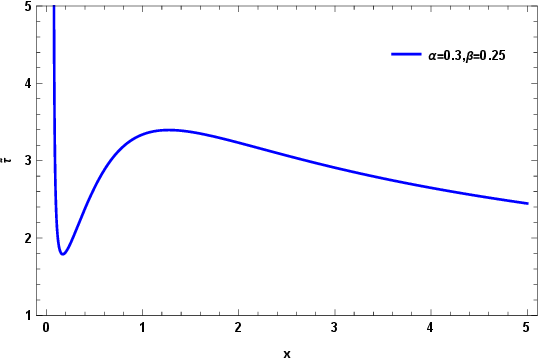}
 \caption{\small{ The plot of the curve of equation \eqref{eq694} with respect to $k=\mathcal{V}=Q=\ell=\Sigma=1$, $y=0.01$, $\alpha=0.3, \beta=0.25$ and $C>C_c=2.5$}}
 \label{fig300}
 \end{center}
 \end{figure}

\begin{figure}[h!]
 \begin{center}
 \subfigure[]{
 \includegraphics[height=4cm,width=5cm]{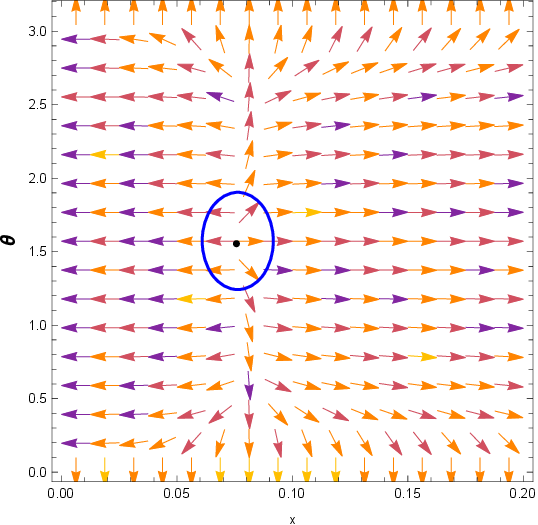}
 \label{fig1a}}
 \subfigure[]{
 \includegraphics[height=4cm,width=5cm]{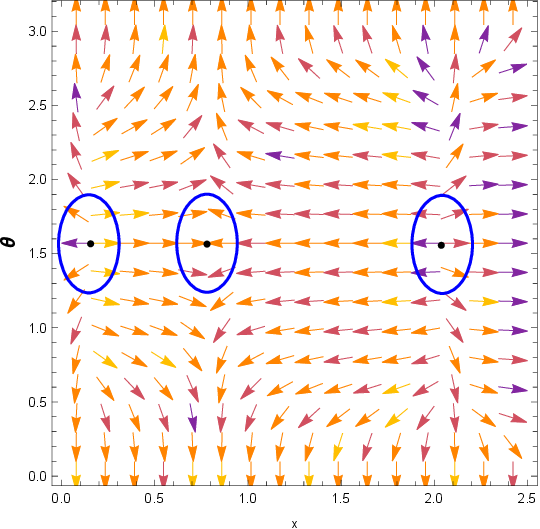}
 \label{fig1b}}
 \subfigure[]{
 \includegraphics[height=4cm,width=5cm]{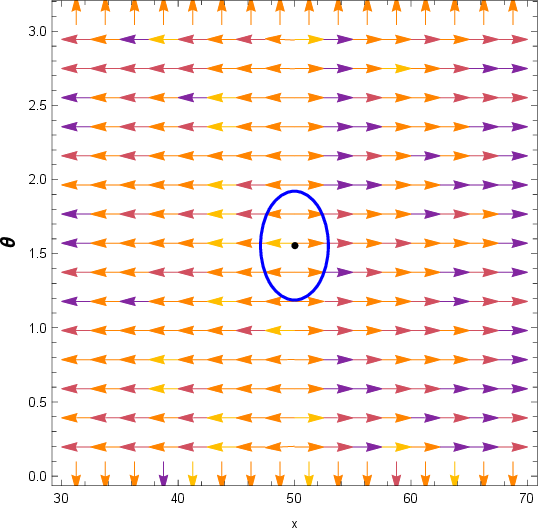}
 \label{fig1b}}
 \caption{\small{the blue arrows represent the vector field n on a portion of the $(x - \theta)$ plane for Einstein-Gauss-Bonnet black holes. The blue loops enclose the ZPs  with respect to $k=\mathcal{V}=Q=\ell=\Sigma=1$, $y=0.01$, $\alpha=0.3, \beta=0.25$ and $C=2.5>C_c=0.187$  (a) $\tilde{\tau}=4$ (b)$\tilde{\tau}=3.2$ , (c)$\tilde{\tau}=1$}}
 \label{fig400}
 \end{center}
 \end{figure}

\begin{figure}[h!]
 \begin{center}
 \subfigure[]{
 \includegraphics[height=4cm,width=5cm]{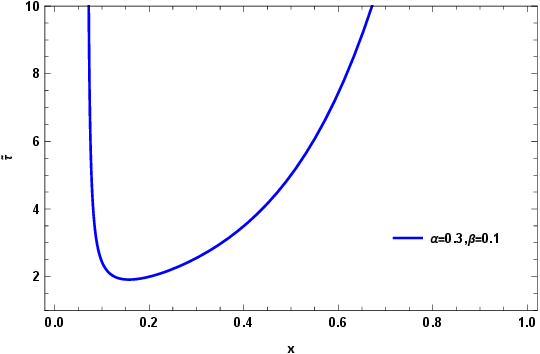}
 \label{fig1a}}
 \subfigure[]{
 \includegraphics[height=4cm,width=5cm]{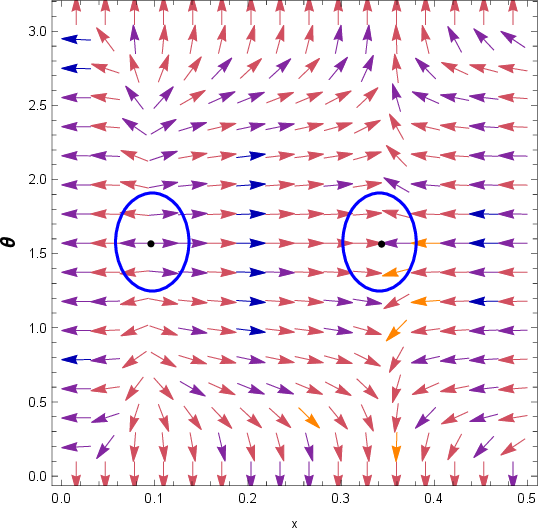}
 \label{fig1b}}
 \caption{\small{ (a)  The plot of the curve of equation\eqref{eq694} with respect to $k=\mathcal{V}=Q=\ell=\Sigma=1$, $y=0.01$, $\alpha=0.3, \beta=0.1$ and $C=2.5>C_c=0.187$ (b), the blue arrows represent the vector field n on a portion of the $(x - \theta)$ plane for Einstein-Gauss-Bonnet black holes. The blue loops enclose the ZPs $\tilde{\tau}=3$ .}}
 \label{fig500}
 \end{center}
 \end{figure}

\begin{figure}[h!]
 \begin{center}
 \subfigure[]{
 \includegraphics[height=4cm,width=5cm]{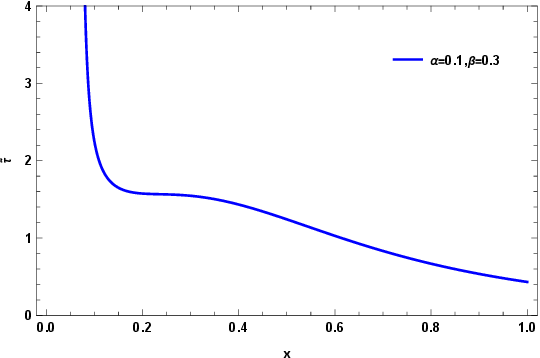}
 \label{fig1a}}
 \subfigure[]{
 \includegraphics[height=4cm,width=5cm]{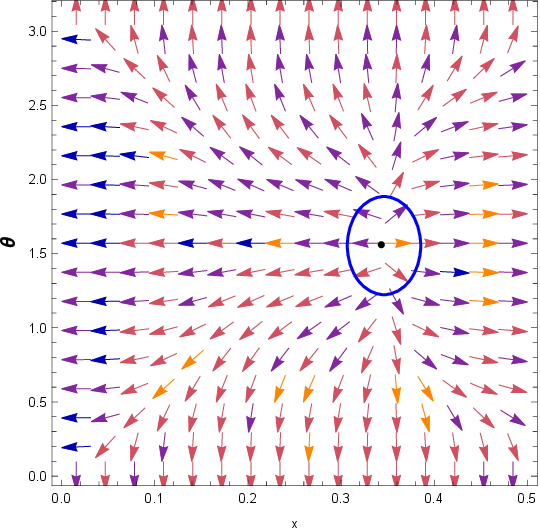}
 \label{fig1b}}
 \caption{\small{ (a)  The plot of the curve of equation \eqref{eq694} with respect to $k=\mathcal{V}=Q=\ell=\Sigma=1$, $y=0.01$, $\alpha=0.1, \beta=0.3$ and $C=2.5>C_c=0.187$ (b), the blue arrows represent the vector field n on a portion of the $(x - \theta)$ plane for Einstein-Gauss-Bonnet black holes. The blue loops enclose the ZPs $\tilde{\tau}=1.5$.}}
 \label{fig600}
 \end{center}
 \end{figure}

In this scenario, a first-order phase transition occurs for the black hole when considering the Bekenstein-Hawking entropy. However, the situation shifts when the Sharma-Mittal entropy is applied.
When $\alpha>\beta$ (Fig. \eqref{fig500}), the total topological charge is 0. When $\alpha$  and $\beta$  are close together ($\alpha\approx\beta$), the topological charge is +1 (Figs. \eqref{fig300},\eqref{fig400} ). Similarly, when $\alpha<\beta$, the topological charge is also +1 (Fig. \eqref{fig600}).
\section{Thermodynamic topology within LQG}
\begin{figure}[h!]
 \begin{center}
 \subfigure[]{
 \includegraphics[height=4cm,width=4cm]{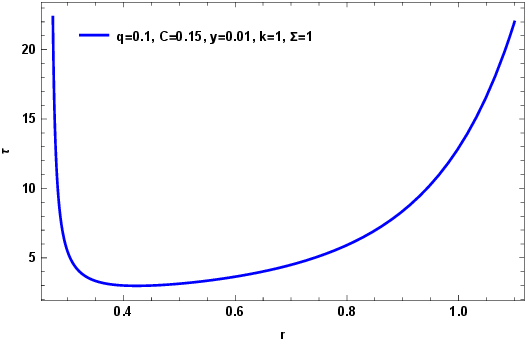}
 \label{600a}}
 \subfigure[]{
 \includegraphics[height=4cm,width=4cm]{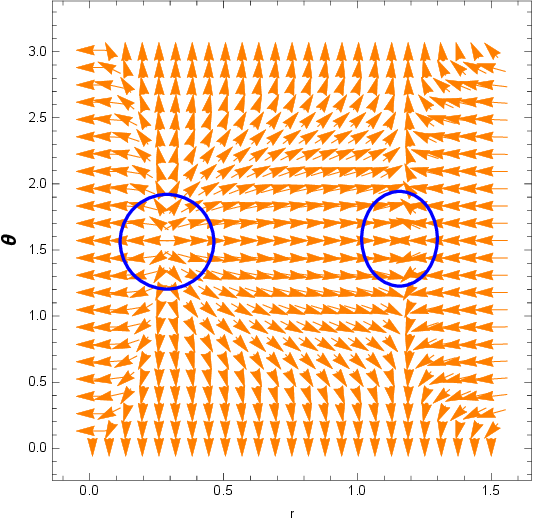}
 \label{600b}}
 \subfigure[]{
 \includegraphics[height=4cm,width=4cm]{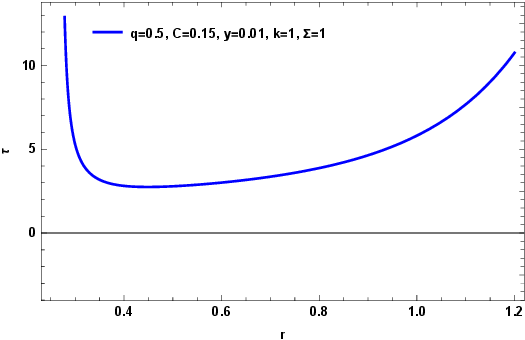}
 \label{600c}}
 \subfigure[]{
 \includegraphics[height=4cm,width=4cm]{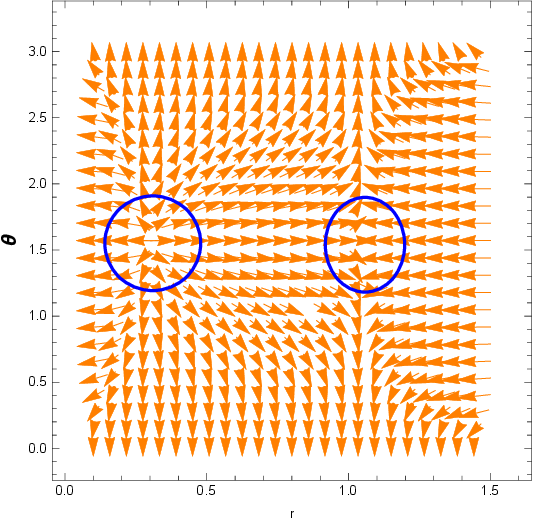}
 \label{600d}}\\
 \subfigure[]{
 \includegraphics[height=4cm,width=4cm]{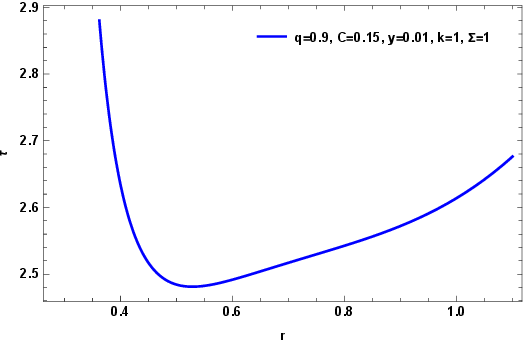}
 \label{600e}}
 \subfigure[]{
 \includegraphics[height=4cm,width=4cm]{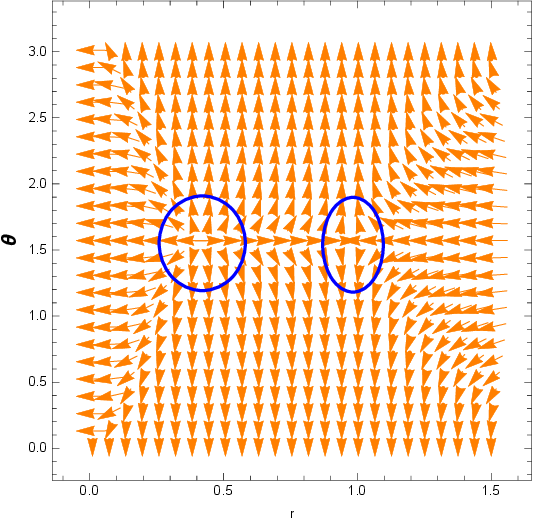}
 \label{600f}}
 \subfigure[]{
 \includegraphics[height=4cm,width=4cm]{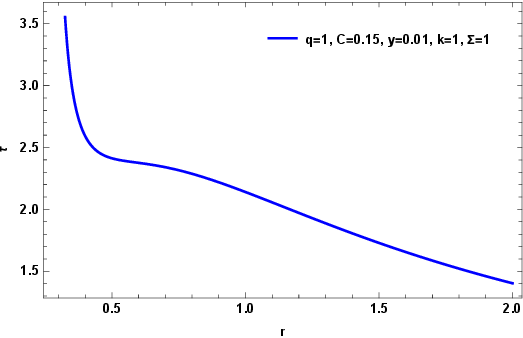}
 \label{600g}}
 \subfigure[]{
 \includegraphics[height=4cm,width=4cm]{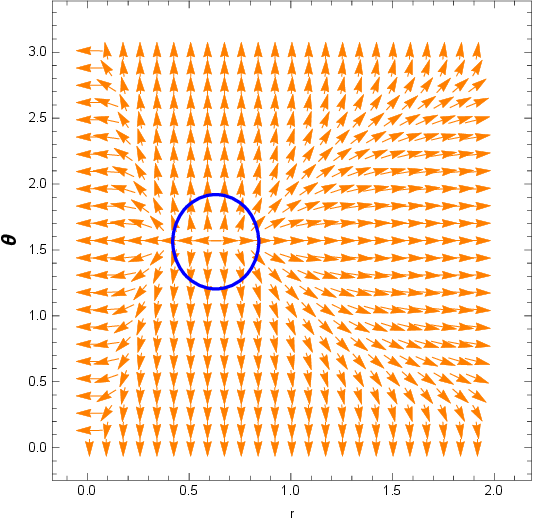}
 \label{600h}}
  \caption{\small{The curve defined by Eq. (\ref{ML4}) is depicted in Figs. (\ref{600a}), (\ref{600c}), (\ref{600e}), and (\ref{600g}). In Figs. (\ref{600b}), (\ref{600d}), (\ref{600f}), and (\ref{600h}), the zero points (ZPs) are positioned at coordinates $(r, \theta)$ which correspond to the free parameters, r=x with $\ell=1$.}}
 \label{m600}
 \end{center}
 \end{figure}

\begin{figure}[h!]
 \begin{center}
 \subfigure[]{
 \includegraphics[height=4cm,width=4cm]{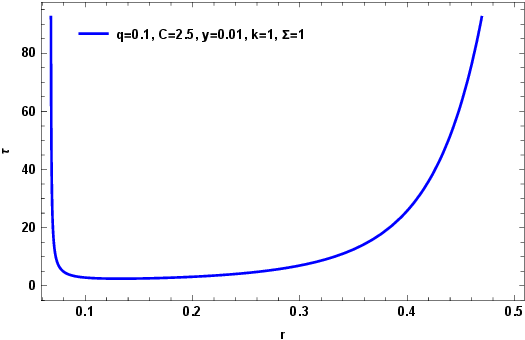}
 \label{700a}}
 \subfigure[]{
 \includegraphics[height=4cm,width=4cm]{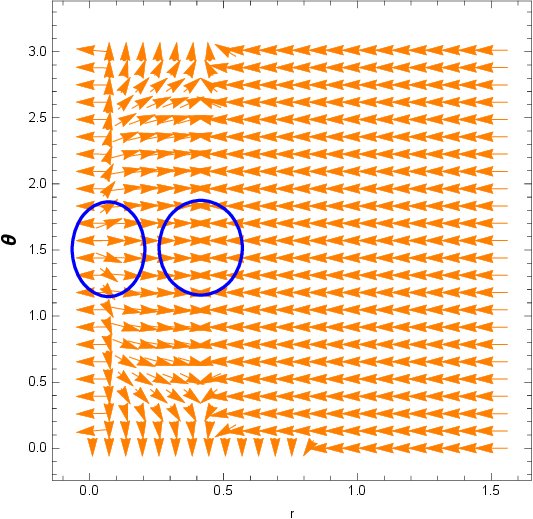}
 \label{700b}}
 \subfigure[]{
 \includegraphics[height=4cm,width=4cm]{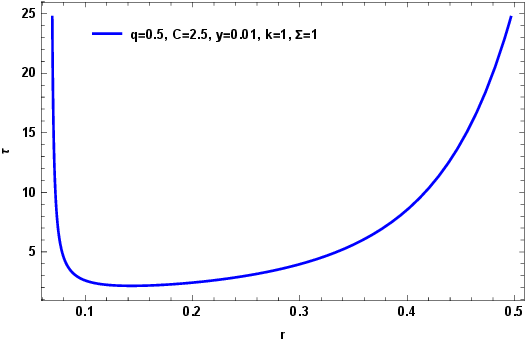}
 \label{700c}}
 \subfigure[]{
 \includegraphics[height=4cm,width=4cm]{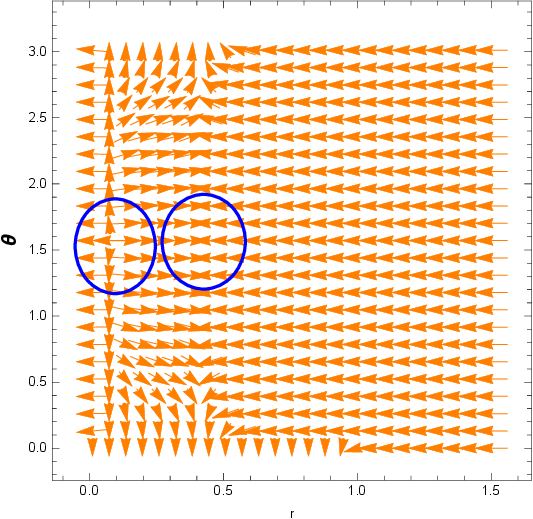}
 \label{700d}}\\
 \subfigure[]{
 \includegraphics[height=4cm,width=4cm]{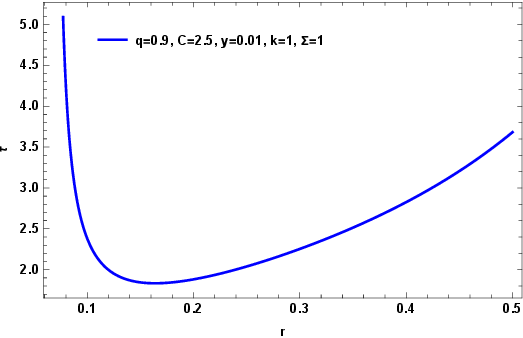}
 \label{700e}}
 \subfigure[]{
 \includegraphics[height=4cm,width=4cm]{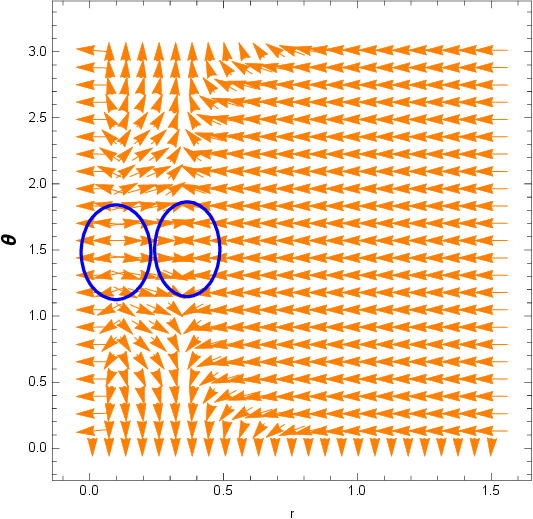}
 \label{700f}}
 \subfigure[]{
 \includegraphics[height=4cm,width=4cm]{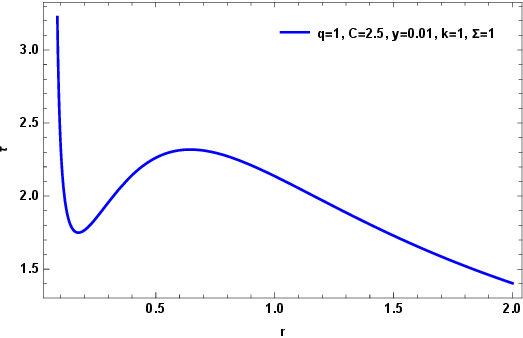}
 \label{700g}}
 \subfigure[]{
 \includegraphics[height=4cm,width=4cm]{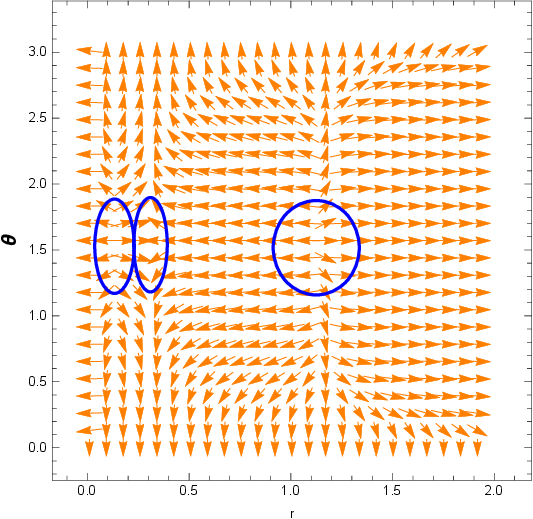}
 \label{700h}}
  \caption{\small{The curve represented by Eq. (\ref{ML4}) is shown in Figs. (\ref{700a}), (\ref{700c}), (\ref{700e}), and (\ref{700g}). In Figs. (\ref{700b}), (\ref{700d}), (\ref{700f}), and (\ref{700h}), the zero points (ZPs) are identified at coordinates $(r, \theta)$ which correspond to the free parameters, r=x with $\ell=1$}}
 \label{m700}
 \end{center}
 \end{figure}

In this study, we investigate the thermodynamic topology of Einstein-Gauss-Bonnet black holes within the context of LQG and the CFT. By utilizing Eqs. (\ref{N3}), (\ref{F1}), we derive the function $\mathcal{F}$.
\begin{equation}\label{ML1}
\mathcal{F}=\frac{\Sigma  \left(768 \pi ^2 c^2 x^6+768 \pi ^2 c^2 x^4+768 \pi ^2 c^2 x^2 y+k^2 Q^2\right)}{256 \pi ^2 c k^{2/3} \sqrt[3]{\mathcal{V}} x^2}-\frac{e^{\frac{4 \pi  c (1-q) \Sigma  \left(x^3+6 x y\right)}{k}}-1}{(1-q) \tilde{\tau}}
\end{equation}
Then, by using Eqs. (\ref{F2}), the $\phi^{r_h}$ and $\phi^{\theta }$ are obtained as follows,
\begin{equation}\label{ML2}
\begin{split}
&\phi^x=\frac{\Sigma  \left(4608 \pi ^2 c^2 x^5+3072 \pi ^2 c^2 x^3+1536 \pi ^2 c^2 x y\right)}{256 \pi ^2 c k^{2/3} \sqrt[3]{\mathcal{V}} x^2}-\frac{\Sigma  \left(768 \pi ^2 c^2 x^6+768 \pi ^2 c^2 x^4+768 \pi ^2 c^2 x^2 y+k^2 Q^2\right)}{128 \pi ^2 c k^{2/3} \sqrt[3]{\mathcal{V}} x^3}\\&-\frac{4 \pi  c \Sigma  \left(3 x^2+6 y\right) e^{\frac{4 \pi  c (1-q) \Sigma  \left(x^3+6 x y\right)}{k}}}{k \tilde{\tau} }
\end{split}
\end{equation}
and
\begin{equation}\label{ML3}
\phi^{\theta }=-\frac{\cot (\theta )}{\sin (\theta )}
\end{equation}
We can calculate the $\tau$ as follows,
\begin{equation}\label{ML4}
\tilde{\tau} =\frac{1536 \pi ^3 \left(2 c^2 \sqrt[3]{v} x^3 y e^{\frac{4 \pi  c (1-q) \Sigma  \left(x^3+6 x y\right)}{k}}+c^2 \sqrt[3]{\mathcal{V}} x^5 e^{\frac{4 \pi  c (1-q) \Sigma  \left(x^3+6 x y\right)}{k}}\right)}{1536 \pi ^2 c^2 \sqrt[3]{k} x^6+768 \pi ^2 c^2 \sqrt[3]{k} x^4-k^{7/3} Q^2}
\end{equation}
Here, we delve into the thermodynamic topology of Einstein-Gauss-Bonnet black holes within the framework of the CFT, taking into account the effects of non-extensive entropy formulations such as LQG. The illustrations are divided, with normalized field lines shown on the right. Figs.(\ref{m600}) and (\ref{m700}) illustrate the results for $C < C_{c}$ and $C > C_{c}$, respectively. Figs. (\ref{600b}), (\ref{600d}), and (\ref{600f}) depict two zero points for $C < C_{c}$, while Figs. (\ref{700b}), (\ref{700d}), and (\ref{700f}) show two zero points for $C > C_{c}$. These zero points, representing topological charges, are determined by the free and non-extensive parameters $q$ and are located within the blue contour loops at coordinates $(r, \theta)$. The sequence of these illustrations is governed by the parameter $q$. The findings from these figures reveal a distinctive feature: two topological charges $(\omega = +1, -1)$ and the total topological charge $W = 0$, indicated by the zero points within the contour. Our analysis evaluates black hole stability by examining the winding numbers. Additionally, as shown in Fig. (\ref{600h}) for $C < C_{c}$ and Fig. (\ref{700h}) for $C > C_{c}$, when the parameter $q$ increases to $1$, the classification changes. We observe one and three topological charges with a total topological charge $W = +1$. Fig. (\ref{600h}) shows one topological charge $(\omega = +1)$ with the total topological charge $W = +1$, while Fig. (\ref{700h}) shows three topological charges $(\omega = +1, -1, +1)$ with the total topological charge $W = +1$.
\subsection{Case I: $C=0.15<C_{c}$}
\subsection{Case II: $C=2.5 > C_{c}$}

\section{Conclusions}

We conducted an in-depth investigation into the thermodynamic topology of Einstein-Gauss-Bonnet black holes within the framework of Conformal Field Theory (CFT), considering the implications of non-extensive entropy formulations. The intriguing results obtained in this study are as follows:

The choice of the parameter $\lambda$ below the critical value (C) has practically no effect on the phase behavior of the black hole. However, when $\lambda$ exceeds the critical value (C), this choice can significantly alter the phase transition outcome of the black hole. Determining which values of $\lambda$ best represent physical reality will undoubtedly require experimental evidence. Nevertheless, this flexibility in choosing the parameter can provide researchers with greater freedom to explain the events occurring during a black hole phase transition, depending on the physical conditions.

When the Sharma-Mittal entropy is incorporated into the Helmholtz free energy to study phase transitions, the parameters $\alpha$ and $\beta$ can influence both the phase transition and the topological charge. Specifically, if $\alpha > \beta$, the topological charge is 0, whereas for $\alpha \leq \beta$, the topological charge is +1. Also, for LQG, As the parameter $q$ increases to 1, the classification changes. We observe one and three topological charges with the total topological charge (W = +1) with respect to $C_{c}$. In other cases, we encounter two topological charges $(\omega = +1, -1)$, leading to a total topological charge (W = 0).

\end{document}